\documentclass[runningheads]{llncs}

\usepackage{eccvabbrv}
\usepackage{amsmath}

\usepackage{graphicx}
\usepackage{booktabs}
\usepackage{algorithm}
\usepackage{algpseudocode}
\usepackage{stackengine}
\usepackage{multirow}
\usepackage[accsupp]{axessibility}  

\usepackage[pagebackref,breaklinks,colorlinks,citecolor=blue]{hyperref}

\usepackage{orcidlink}

\stackMath

\begin{document}

\title{ $DLOVE$: A new Security Evaluation Tool for Deep Learning Based Watermarking Techniques}

\author{Sudev Kumar Padhi\\
Indian Institute of Technology, Bhilai\\
Durg, Chattisgarh, 491002\\
{\tt\small sudevp@iitbhilai.ac.in}
\and
Dr. Sk. Subidh Ali\\
Indian Institute of Technology, Bhilai\\
Durg, Chattisgarh, 491002\\
{\tt\small subidh@iitbhilai.ac.in}
}

\maketitle
\begin{abstract}
  Recent developments in Deep Neural Network ($DNN$) based watermarking techniques have shown remarkable performance.  The state-of-the-art $DNN$-based techniques not only surpass the robustness of classical watermarking techniques but also show their robustness against many image manipulation techniques. In this paper, we performed a detailed security analysis of different $DNN$-based watermarking techniques. We propose a new class of attack called the Deep Learning-based OVErwriting ($DLOVE$) attack, which leverages adversarial machine learning and overwrites the original embedded watermark with a targeted watermark in a watermarked image. To the best of our knowledge, this attack is the first of its kind. To show adaptability and efficiency, we launch our $DLOVE$ attack analysis on four different watermarking techniques, $HiDDeN$, $ReDMark$, $PIMoG$, and  {\em Hiding Images in an Image}. All these techniques use different approaches to create imperceptible watermarked images. Our attack analysis on these watermarking techniques with various constraints highlights the vulnerabilities of $DNN$-based watermarking. Extensive experimental results validate the capabilities of $DLOVE$. We propose $DLOVE$ as a benchmark security analysis tool to test the robustness of future deep learning-based watermarking techniques.

\keywords{Deep Learning \and  Adversarial Machine Learning ($AML$) \and Digital Watermarking.}

\end{abstract}
\section{Introduction}
Digital watermarking is a well-known technique where the watermark (message or image) is embedded covertly or overtly into a cover image without distorting the quality of the cover image~\cite{survey2,cox2002digital,cox2002first,shaik2022review,Coxwatermark}. It has various critical applications, such as copyright protection, content authentication, tamper detection, data hiding, etc.  In watermarking, the sender embeds the watermark into the cover image and sends the watermarked image to the receiver or verifier. To validate the authenticity or copyright, the watermark from the received watermarked image is extracted and compared with the original watermark, which is provided to the receiver or verifier in advance. Generally, watermarking techniques consist of two processes: watermark embedding and watermark extraction. In watermark embedding, the watermark is embedded into the input cover image to produce a watermarked image. While in the watermark extraction process, the watermark is extracted from the watermarked image and compared with the original watermark to validate the ownership or authenticity of the cover image.  One of the popular watermarking techniques is invisible watermarking, where the watermark is covertly embedded in the cover image. The security of any invisible watermarking techniques lies in the secrecy of the embedded watermark, such that the watermarked image should be perceptually similar to the cover image and should not contain any detectable artifact.  

The classical watermarking techniques use a wide variety of embedding approaches from the spatial and frequency domains~\cite{4303093,raj2021survey,Wong1998APK,ping,wang2002wavelet,robust,robust1,951542}. Recently, deep learning has emerged as the key enabler of $AI$ applications. Thus, there has been an increase in deep learning techniques using Deep Neural Networks ($DNN$) for different tasks due to their adaptability in various applications. It is also being utilized in the domain of watermarking techniques, which has resulted in significant improvements in performance and efficiency compared to traditional techniques~\cite{ref4}. In $DNN$-based watermarking techniques,  the watermark embedding and extraction processes are implemented using deep generative networks, such as autoencoders and Generative Adversarial Networks ($GAN$). The pioneering $DNN$-based watermarking technique proposed in~\cite{ref4} can hide an $RGB$ image within another $RGB$ image using an autoencoder network. $DNN$-based watermarking was further enhanced by introducing distortion into the training data to make the watermarked images robust against certain noises~\cite{lacuna5,deepwatermark1}. These simple autoencoder-based techniques are vulnerable to Deep Learning based Removal ($DLR$) attacks~\cite{pixel,dest,liu2023erase}. There are different types of $DLR$ attacks. In one of the approaches, the attacker trains a denoising autoencoder to remove the watermark from the watermarked image as noise~\cite{dest}. In another approach, the pixel distribution of the watermarked image is used to identify the distorted pixels for removing the watermark~\cite{pixel}. Pixel impainting technique is also utilized to remove the watermark from the watermarked image~\cite{liu2023erase}. In this line, the watermarking technique proposed in~\cite{hiddn,deepwatermark2,lacuna6,lacuna1} is considered to be robust against $DLR$ attacks due to the presence of noise layers in their model architectures. Among these, the most popular technique is $HiDDeN$~\cite{hiddn}, which can withstand arbitrary types of image distortion and makes robust watermarked images. $PIMoG$~\cite{fang2022pimog} went one step ahead by introducing screen-shooting robustness such that the watermark can be extracted even if the digital image is captured with a camera. This robustness is achieved by introducing a mask-guided loss in the training pipeline of the watermarking technique. Similarly, $ReDMark$~\cite{lacuna4}  uses residual structure to embed the watermark, striking a balance between robustness and impermeability.

Please note that $DLR$ attacks are useful for limited applications where the attacker's objective is just to fail the ownership claim of the actual owner of the cover image. The attacker cannot claim ownership of the cover image using $DLR$ attacks. In order to claim ownership of a cover image, the attacker has to overwrite the original watermark of a given watermarked image with the attacker's watermark such that the watermark extraction process should extract the attacker's watermark from the watermarked image instead of the original watermark. There is no doubt that classical watermark overwriting attacks will not work on $DNN$-based watermarking technique techniques~\cite{qian2015deep,pibre2015deep}. It requires a Deep Learning based OVErwriting ($DLOVE$) attack. However, there is hardly any work in the open literature related to the $DLOVE$ attack. In regular deep learning applications, similar attacks are common, which are known as Adversarial Machine Learning ($AML$) attacks~\cite{intriguing,harness,papernot2016limitations,nguyen2015deep}. In targeted $AML$, the attacker induces a well-crafted perturbation into the input image such that the model used for classification not only fails to classify it but is also forced to misclassify it into a target class as desired by the attacker. We can intuitively consider that the attacker is overwriting the features of the original class in the input image with the features of the target class.  Inspired by targeted $AML$ attacks, for the first time, we developed the $DLOVE$ attack against $DNN$-based watermarking techniques.

In this paper, we perform a security analysis of $DNN$-based watermarking techniques using the $DLOVE$ attack. Here, the robustness of these $DNN$-based watermarking techniques is verified against well-crafted perturbations where the final goal is to overwrite the embedded watermark with the desired watermark. The attack is targeted for the real-world scenario where the watermarking techniques are used to perform copyright protection. To show the adaptability and efficiency, we launch our $DLOVE$ attack on four different watermarking techniques, which are  $HiDDeN$, $ReDMark$~\cite{lacuna4}, $PIMoG$~\cite{fang2022pimog}, and {\em Hiding Images in an Image}~\cite{ref4}. All these techniques use different approaches to create imperceptible watermarked images. Devising a common approach to attack these techniques with various constraints highlights the vulnerabilities of $DNN$-based watermarking.


The paper makes the following key contributions:
\begin{enumerate}
    \item  We are the first to propose $DLOVE$, a watermarking overwriting attack based on the concept of targeted $AML$ to overwrite the embedded watermark with the target watermark by adding well-crafted perturbation to the watermarked images.
    \item  We introduce a new class of attack solely using the knowledge available when $DNN$-based watermarking techniques are used for copyright protection.
    \item A detailed experimental result is provided to validate the success of the $DLOVE$ attack. The results demonstrate that the $DLOVE$ attack generalizes well on different $DNN$-based watermarking techniques.

\end{enumerate}

\section{ Related Works}

\subsection{Deep learning based Watermarking}

 Recently, many $DNN$-based watermarking techniques have been proposed, which surpass the performance of traditional watermarking by utilizing the efficient feature extraction ability of the neural networks.  The main architecture used in $DNN$-based watermarking involves the use of an encoder network that embeds the watermark into the cover image and a decoder network that extracts the watermark from the watermarked image. $DNN$-based watermarking can embed an image or bit string as a watermark but most techniques choose to embed a bit strings. Bit strings work as metadata and provide more robustness compared to embedding images as a watermark. This is due to the fact that embedding an image requires the decoder to learn the spatial information of the watermark, which can hamper robustness. The training of the encoder and decoder is done in an end-to-end manner as a pipeline~\cite{ref4,deepwatermark1,fang2020deep}. To further enhance the quality and robustness of the watermarked image, a discriminator is added in the pipeline while training and noise layers are added in the model architecture of the $DNN$-based watermarking~\cite{hiddn,lacuna1,fang2022pimog}. The discriminator acts as an adversary network, which predicts whether the watermark is embedded in an image. Residual connections and layers of random combinations of a fixed set of distortions are also used in some model architectures to make the watermarking technique more robust with high data hiding capacity~\cite{lacuna4,lacuna2}. Almost all of these $DNN$-based methods achieve great performance in terms of image quality. Generally, when we consider robustness in watermarking, it refers to handling distortion that exists in image processing, such as $JPEG$ compression, blurring, noises, crop out, etc. There is hardly any analysis that aims to find the vulnerability of $DNN$-based watermarking techniques against the $DLOVE$ attacks.

\subsection{Adversarial Machine Learning}
$AML$ attacks have the capability to fail highly accurate machine learning models~\cite{intriguing,harness,papernot2016limitations,nguyen2015deep}
by adding a well-crafted perturbation into the input image. These attacks are majorly developed to fail deep convolution neural network-based classifiers. Transferable $AML$ attacks are also developed~\cite{liu2016delving,zhou2018transferable,papernot2016transferability,papernot2017practical} such that a perturbation crafted to fail one model can also be used to fail other models that perform a similar task even if the attacker has no access to the second model's parameters or architecture. In $AML$, knowledge of the attacker is assumed to be either white-box (complete knowledge of the target model architecture, its parameters and training data)~\cite{harness,towards,deepfool} or black-box access ( limited or no knowledge to the target model)~\cite{simba,limba}. In a white-box attack, the attacker can craft adversarial examples by directly manipulating the input data to maximize the model's loss or misclassification using the model parameters. While in a black-box attack, despite lacking internal knowledge of the model, the attackers can still generate adversarial examples by exploiting the model's response to input queries. 
These queries can be carefully chosen such that by observing its outputs, information about the model can be inferred, and adversarial examples can be crafted accordingly using the transferability of $AML$ attacks.

$AML$ is a great tool for the designer of deep convolution neural network-based classifiers to test the robustness of their classifiers. There is a lack of such tools in the domain of $DNN$-based watermarking techniques. In this paper, we tried to overcome this lacuna by introducing the $DLOVE$ attack, which can be an interesting tool for the designers of $DNN$-based watermarking techniques. Our approach is inspired by targeted $AML$ attacks.  The objective of the $DLOVE$ attack is to craft a new watermark, which, once added to the watermarked image, will force the watermark decoder to decode the new watermark instead of the original watermark. We demonstrate the $DLOVE$ attack in the white box as well as black box settings.

\section{Threat Model}

\subsection{Attackers Goals}
Copyright protection is one of the most important use cases of watermarking through which the owner of digital content can claim its rights.  An attacker can violate the copyright protection either by corrupting/cleaning the embedded watermark in the image so that the decoder cannot decode the watermark from the watermarked image ({\bf Objective 1}) or by overwriting the embedded watermark present in the watermarked image with the target watermark so that the decoder will decode the target watermark instead of the original watermark from the watermarked image ({\bf Objective 2} ). In either case, the objective is to defeat the techniques of watermarking. 

\begin{figure}[!ht]
\centering
\includegraphics[width=\linewidth]{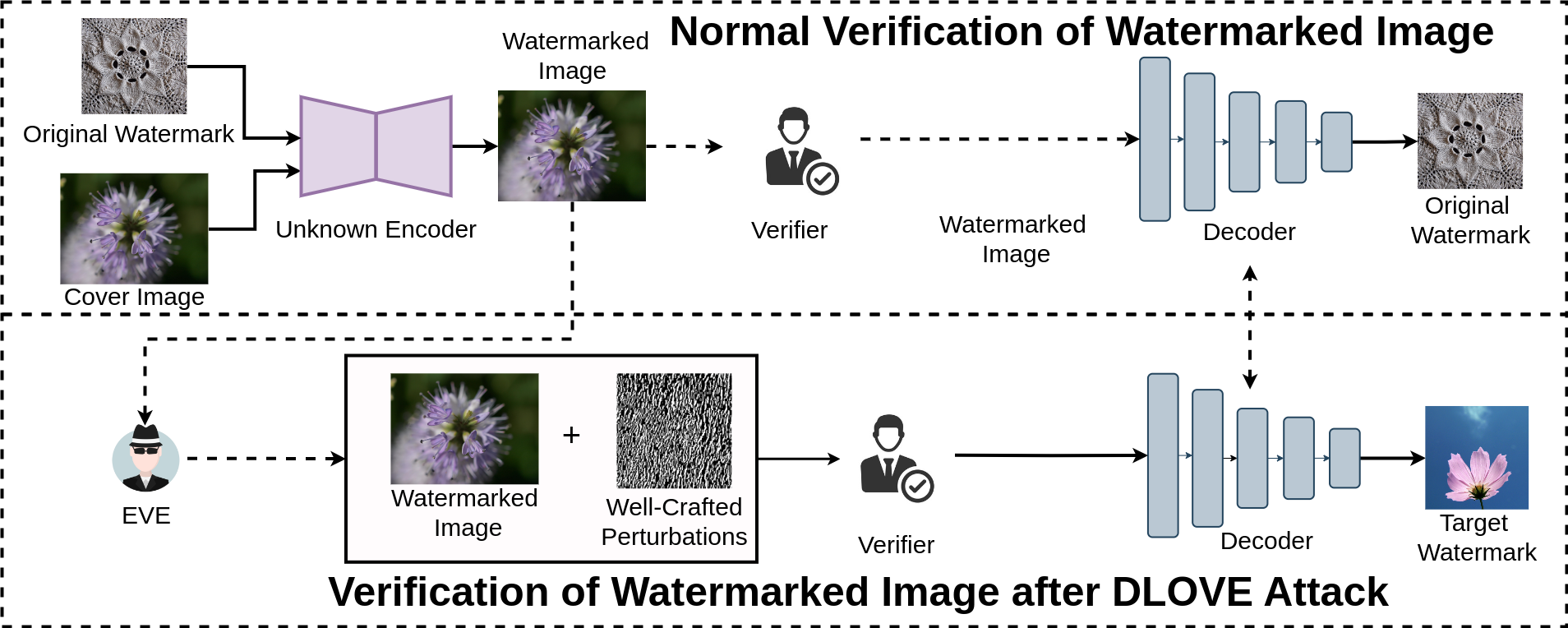}
\caption{Overview of the proposed $DLOVE$ attack leveraging Adversarial Machine Learning to a create well-crafted perturbation to overwrite the original watermark with the target watermark.}
\label{Model}
\end{figure}

\vspace{-2em}

\subsection{Attackers Knowledge}  

 Before going into the details of the attack, we make the following assumptions about the attacker's knowledge:
\\
\noindent \textbf{ Training Data:}
The attacker has no knowledge of the training data used to train the $DNN$-based  watermarking model in both variants of the $DLOVE$ attack (white box and black box settings). This included both the watermark and the cover image. 
\\
\noindent \textbf{Network Architecture:}   The architecture of the encoder network is not known to the attacker in both black and white box settings. In the white-box variant of the attack, it is assumed that the attacker has knowledge of the decoder network architecture and its parameters.  The same is not valid for the black-box setting of the $DLOVE$ attack.

The black box setting of the $DLOVE$ attack is more practical and useful in professional applications of watermark~\cite{realapp1,realapp2,realapp3,realapp4}, where the watermarking technique is available as a service ($API$)  to verify the digital content. In such a scenario,  the attacker can subscribe to the service and get Oracle access to both the encoder and decoder through its $API$. Nevertheless, there is a limit on the number query to the $API$. However, in stringent secure application scenarios, even Oracle access to the decoder is infeasible for the attacker as it remains under the possession of the verifier only. 
The $DLOVE$ attack considers these stringent security assumptions in the black-box setting.

\subsection{Scenario}

Let $Alice$ be a digital artist who creates digital paintings. She wants to protect her digital paintings (copyright) from unauthorized use and distribution. $Alice$ uses $DNN$-based invisible watermarking as it protects the copyright of the painting and also preserves its aesthetic appeal. The watermarking technique subscribed by Alice uses the logos of the artist as the watermark. Thus, Alice embeds the logo of her website into her digital paintings (cover image). For verification, the verifier needs to find the presence of a watermark, extract it, and verify the owner of the digital painting. In copyright protection, similar information that forms the metadata of the digital content for different owners is used as a watermark (in this case, it is a logo). This is to make sure that the verifier can verify with consistent information. Now, there is an attacker, $Eve$, who has also subscribed to the same watermarking technique used by Alice. Thus, she knows that the digital paintings of Alice are copyright-protected with the logo of Alice's website.   $Eve$ can clean or overwrite the watermark with a target watermark containing a different logo in the watermarked image and recirculate it. By achieving {\em Objective 1}, $Eve$ can only remove the watermark from the digital painting. While achieving {\em Objective 2}, $Eve$ not only removes the watermark but also makes herself the digital painting owner by embedding her logo into it. $Alice$ cannot prove that the digital painting belongs to her as the decoder decodes the logo, which belongs to $Eve$. This scenario is depicted in Figure~\ref{Model}, where the decoder decodes the target watermark instead of the original watermark when the well-crafted perturbation is added to the watermarked image. Thus, the verifier will announce that the digital paintings belong to $Eve$.

\section{Proposed Approach}
\label{approach}
\subsection{Formal description}
$DNN$-based watermarking techniques consist of an encoder and a decoder. The encoder $E$ produces a watermarked image $W$ by embedding the watermark   $\alpha$ into the cover image $I$ as shown in Eq.~\eqref{eq1}. In contrast, the decoder $D$ takes $W$ as input and extracts the embedded watermark  $\alpha$ as the output, as shown in Eq.~\eqref{eq2}.  The attacker's aim is to launch the $DLOVE$ attack to fool $D$ by inducing adversarial perturbation $\delta$ in $W$ such that $D$ decodes the target watermark  $\beta$ instead of the original embedded watermark $\alpha$ as shown in Eq.~\eqref{eq22}. 

\begin{eqnarray}
    E(I + \alpha) \rightarrow W  \label{eq1}\\
   D(W) \rightarrow \alpha \label{eq2}\\
   D(W+\delta)\rightarrow \beta \label{eq22}
\end{eqnarray}

\subsubsection{White-Box Access:} Having white-box access to the decoder gives the attacker enough information to simulate the network by devising a targetted adversarial attack and using the gradients of the decoder to create the desired perturbation $\delta$, where $\alpha$ is the original watermark, $\beta$ is the target watermark and $\epsilon$ is the perturbation limit.  We minimize the loss ($l$) of  $\beta$, which is the target watermark while maximizing the loss of $\alpha$, which is the original watermark, i.e. we solve the optimization problem as shown in Eq.~\eqref{eq3}. This is the easiest approach but does not align with the use cases of watermarking, where access to the decoder is not allowed.  

\begin{equation}
        \stackunder{minimize} {\delta} \{l( D(W + \delta ), \beta ) - l(D(W + \delta ), \alpha)\} , \quad \delta \in [-\epsilon,\epsilon]
         \label{eq3}
\end{equation}

\begin{figure}[!htb]
\centering
\includegraphics[width=\linewidth]{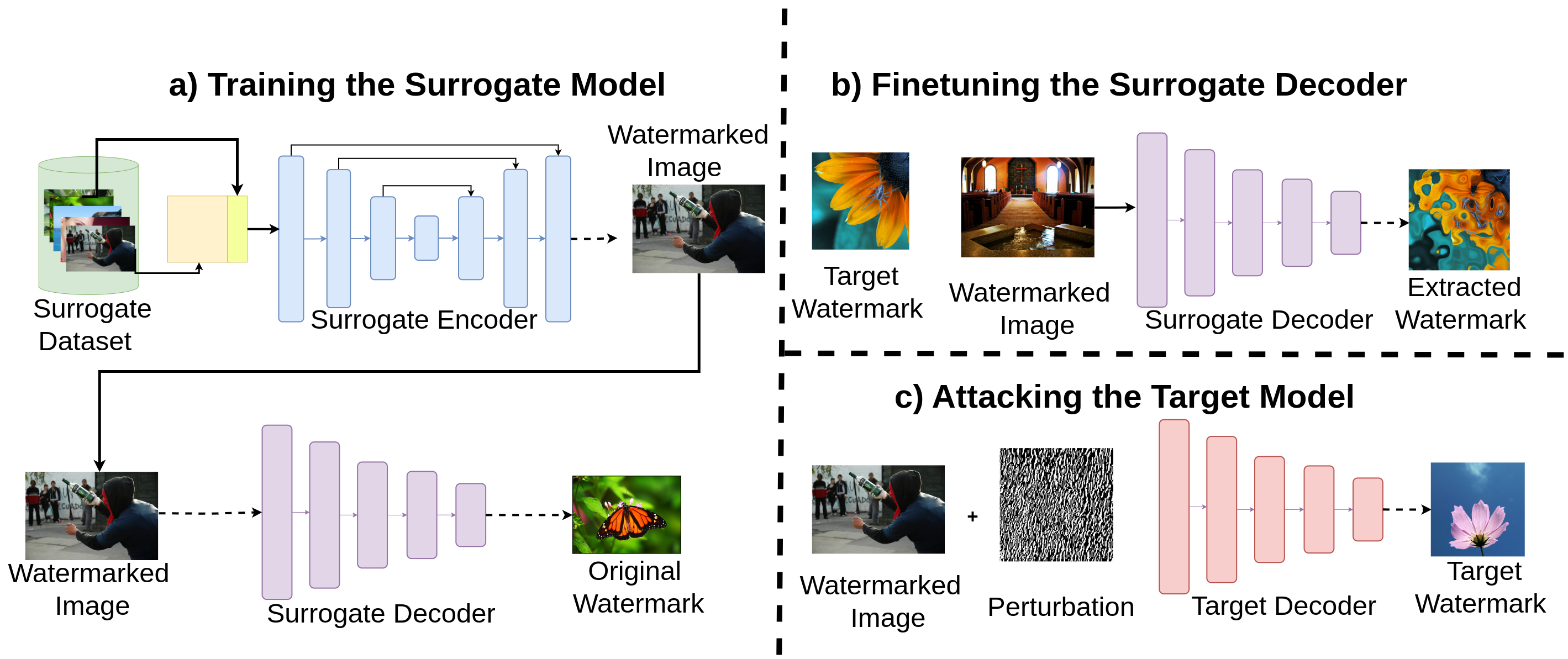}
\caption{Overview of surrogate model attack \textbf{a)} Training the surrogate model using surrogate dataset \textbf{b)} Fine-tuning the surrogate decoder with the watermarked image of the target $DNN$-based 
 watermarking technique \textbf{c)}  Attacking the decoder of the target $DNN$-based watermarking technique after generating the well-crafted perturbation from the surrogate decoder.}
\label{overall}
\end{figure}

\subsubsection{Black-Box Access:} If the attacker has the ability to use the decoder as an oracle, it can obtain a set of watermarked images and their watermarks by querying the decoder with watermarked images.  Once this data set is available, the attacker can train a surrogate decoder. Afterwards, a white-box attack is performed on the surrogate decoder to craft the desired perturbation $\delta$, which is used to launch the $DOVE$ attack on $D$. However, in stringent security applications, even the decoder is not available. Therefore, we consider only having limited instances of watermarked images whose watermarks are known. One of the easiest ways for the attacker to gain access to such data is to request the subscribed copyright protection service provider to copyright on, say, $n$ pairs of cover images and their watermarks. 

Under this scenario, the attacker will train its own $DNN$-based surrogate watermarking encoder ($E^\prime$) and decoder ($D^\prime$) models with its own dataset (also known as the surrogate dataset) $i.e.$, a set of cover images and their watermarks. Once the surrogate model is trained, $D^\prime$ is fine-tuned with the limited instances of watermarked images available from the target decoder $D$ to be attacked. While fine-tuning, loss between the extracted watermark and the original watermark is used for the training of the surrogate decoder, as shown in Eq~\eqref{eq6}.   The surrogate decoder is trained and fine-tuned to act as the target decoder $D$, making the black-box attack transferable. Therefore, the attacker can launch a white-box $DLOVE$ attack on $D^\prime$ using the gradient information to craft the desired perturbation $\delta$ as shown in Eq.~\eqref{eq4}. The same $\delta$ can be used to fail $D$ when added with $W$ (Eq.~\eqref{eq5}).  The value $\epsilon$ is chosen judiciously such that the induced perturbation ($\delta$) to the watermarked image is imperceptible. Figure~\ref{overall} refers to the training procedure of the surrogate model and fine-tuning of the surrogate decoder to perform an attack on the target decoder.

\begin{equation}
   {minimize} \{l(D'(W),\alpha)\}
   \label{eq6}
\end{equation}

\begin{equation}
        \stackunder{minimize} {\delta} \{l( D^\prime(W + \delta ), \beta )\} , \quad \delta \in [-\epsilon,\epsilon]
         \label{eq4}
\end{equation}
\begin{equation} 
        \ D(W + \delta ) \rightarrow \beta\ 
         \label{eq5}
\end{equation}

\subsection{Crafting algorithm}

The adversarial perturbation crafting algorithm is shown in Algo~\ref{algo-crafting-adversarial-samples}. The algorithm shows how to craft a perturbation using the $DLOVE$ attack in a white box scenario. Inputs to the algorithm are, a watermarked image $W$, the target decoder $D$, the target watermark $\beta$, a perturbation $\delta$ (initialized as zero) with the same size ($k\times k$) as $W$,  a limiting range $\epsilon$ of $\delta$ (-$\epsilon$ $\leq$ $\delta$ $\leq$ $\epsilon$).  $\delta$ is added with $W$ and passed into the decoder $D$, which decodes the secret as $\gamma$. The loss between $\gamma$ and $\beta$ is computed using the chosen loss function $l$. In each iteration of the loop, the optimizer tries to minimize the loss between $\gamma$ and $\beta$ and maximize the loss between $\beta$ and $\alpha$.  Accordingly, $\Delta$ is updated. This process is repeated until the model converges and the desired $\delta$ is obtained, which is the realization of the $DLOVE$ attack on $W$ to overwrite $\alpha$ with $\beta$.

\begin{algorithm}
\caption{\textbf{Adversarial perturbation crafting algorithm }
}
\label{algo-crafting-adversarial-samples}
\begin{algorithmic}[1]
\Require $W$, ${D}$, $\mathbf{\beta}$, $\delta$, $\epsilon$
\State $\delta \gets [0]_{k \times k}$  \Comment{Initial perturbation}
\State max\_iter=$k\times k$ 
\State $\gamma$ = $D$($W + \delta$)\Comment{Embedded Watermark}
\While{$i\leq max\_iter$ \&  $ \gamma \neq$ $\beta$ }
    \State $\gamma\gets D(W + \delta)$  \Comment{Intermediate decoder's output}
    \State $\Delta \gets l(\beta, \gamma) - l (\beta, \alpha)$  
    \State Update $ \delta: \delta_i \gets \delta_{i-1} - \eta \nabla \Delta(\delta)$  \Comment{ Update delta with respect to the loss}
    \State $\delta \gets Clip (\delta,[-\epsilon,\epsilon])$  \Comment{ $\delta$ is clipped}
\EndWhile
\State return $\delta$
\end{algorithmic}
\end{algorithm}

The hyperparameters (parameters that we explicitly define) for this attack include: 
\begin{enumerate}
    \item $\epsilon$: The maximum amount of allowable perturbation that can be added to the images.
    \item Optimizer:  It is used to find the well-crafted perturbation $\delta$.
    \item $l$: The loss function chosen for minimizes the loss between $\gamma$ and $\beta$ and maximizes the loss between $\beta$ and $\alpha$.
\end{enumerate}

This algorithm works similarly in the black-box scenario where the decoder $D$ is replaced by a surrogate decoder $D^\prime$, and the corresponding loss will be $l(\beta, \gamma)$ + $l(D^\prime(W),\alpha)$ in line $6$ of the algorithm. 

\subsection{Reason For Successful Attack}

In the classification task,  whatever may be the input, the classifier will always classify it into one of the classes. These classes are also known to the attacker while performing a targeted adversarial attack. Thus, the attacker adds perturbation such that the boundary of the current class is crossed to the target class and the confidence level of the classifier corresponding to the target class is the highest. Suppose the watermark is of $N$-bit, $i.e.$, the output of the decoder is a $N$-bit watermark. Now, we can consider the decoder as a classifier that classifies the watermarked image into one of $2^{N}$ possible watermark classes. Therefore,  our attack can be considered a targeted adversarial attack where the target class is the target watermark among one of the $2^{N}$ possible cases. $DNN$-based watermarking techniques are trained end-to-end based on the perceptual similarity of the image after embedding a watermark, which makes the embedding region-specific and susceptible to attack. Even if the models are trained for robustness against prominent image manipulation attacks, the same factor is responsible for the generation of adversarial perturbations.

\section{Experimental Results}

In this section, we validate the effectiveness of our $DLOVE$ attack on four well known $DNN$-based watermarking techniques: $HiDDeN$~\cite{hiddn}, $ReDMark$~\cite{lacuna4}, $PIMoG$~\cite{fang2022pimog} and {\em Hiding Images in an Image}~\cite{ref4}. Experiments are conducted on a machine with  $14$-$core$ $Intel$ $i9$ $10940X$ $CPU$, $128$ $GB$ $RAM$, and two $Nvidia$ $RTX$-$5000$ $GPU$s with $16$ $GB$ $VRAM$ each.

\subsection{ Setup of Target Models}
The target $DNN$-based watermarking techniques have different $DNN$ model architectures, training pipelines, training datasets, and watermark sizes. The key features of these four techniques are described below with the help of Table~\ref{table:comp}:
\begin{enumerate}
    \item $HiDDeN$ is an end-to-end model for image watermarking that is robust to arbitrary types of image distortion. It comprises four main components: an encoder, a parameterless noise layer $N$, a decoder, and an adversarial discriminator. The encoder uses a $128$$\times$$128$$\times$$3$ cover image to embed a $30$-bit binary watermark. 
    \item $ReDMark$ uses residual connections, circular convolution, attack layer (simulated attacks during training against real-world manipulations, particularly JPEG compression), and $1$d convolution layers for embedding and extracting the watermark. It takes a grayscale $32$$\times$$32$$\times$$1$ cover image and embeds a $4$$\times$$4$-bit watermark using the residual connection between the layers. 
    \item $PIMoG$ consists of three main parts: the encoder, the screen-shooting noise layer, and the decoder. In order to achieve both screen-shooting robustness by handling perspective distortion, illumination distortion and moi${r}$e distortion while maintaining high visual quality,  the technique uses an adversary network with edge mask-guided image loss and gradient mask-guided image loss. It uses a $128$$\times$$128$$\times$$3$ size cover image to embed a $30$-bit watermark.
    \item {\em Hiding Images in an Image} could hide a $200$$\times$$200$$\times$$3$ image as a watermark inside a $200$$\times$$200$$\times$$3$ cover image. In this technique, the watermarked image is passed through a preparation network that transforms and concatenates it to the original cover image using a hiding network. During decoding, the watermarked image is passed through a reveal network and outputs the watermarked image. 
\end{enumerate}

\vspace{-1.5 em}

\begin{table}[!htb]
\caption{ Characteristics of different $DNN$-based watermarking techniques which are attacked by $DLOVE$. }
\begin{tabular}{|c|c|c|c|c|}
\hline
Technique    & \begin{tabular}[c]{@{}c@{}}Dicrimator \\ in the Loop\end{tabular} & \begin{tabular}[c]{@{}c@{}}Cover \\ Image Size\end{tabular} & \begin{tabular}[c]{@{}c@{}}Watermark\\ Size\end{tabular}           & Dataset  \\ \hline
$HiDDeN$~\cite{hiddn}      & Yes                                                               & $128$$\times$$128$$\times$$3$                                                  & $30$ bit                                                           & $COCO$~\cite{lin2014microsoft}     \\ \hline
$ReDMark$~\cite{lacuna4}      & No                                                                & $32$$\times$$32$$\times$$1$                                                    & $4$$\times$$4$ bit                                                        & $CIFAR10$~\cite{cifar}   \\ \hline
$PIMoG$~\cite{fang2022pimog}        & Yes                                                               & $128$$\times$$128$$\times$$3$                                                  & $30$ bit                                                       & $COCO$~\cite{lin2014microsoft}    \\ \hline
{\em Hiding Images in an Image}~\cite{ref4} & No                                                                & $200$$\times$$200$$\times$$3$                                                  & $200$$\times$$200$$\times$$3$ bit                                               & $ImageNet$~\cite{deng2009imagenet} \\ \hline
\end{tabular}

\label{table:comp}
\end{table}
 
\vspace{-2em}

\subsection{Training Surrogate Model}

Each of the techniques mentioned above uses a different resolution of the cover image and watermark sizes as shown in Table~\ref{table:comp}. Therefore, we have made four instances of the surrogate model (encoder and decoder), $i.e.$, one instance for each target model. The size of the cover image, watermarked image and watermark for each instance of the surrogate model is set according to the target model it corresponds.   We have used $UNet$~\cite{unet} architecture for the surrogate encoder, whereas for the surrogate decoder, after trying various models, we have used two different architectures: one is spatial transformer~\cite{spatial} with seven convolutional layers followed by two fully connected layers and the other is Self-supervised vision transformer ($SiT$)~\cite{sit} based autoencoder. The first one is used to build surrogate decoders for $HiDDeN$, $ReDMark$, and $PIMoG$, where a bit-string is used as the watermark. The second architecture is used to build the surrogate decoder for {\em Hiding Images in an Image}, where an image is used as the watermark.  We have used the $Mirflickr$~\cite{mir} dataset as our surrogate dataset, consisting of one million images with varied contexts, lighting, and themes, from the social photography site $Flickr$. We used $50$k images in our training set and $10$k in our test set. We trained all four surrogate models in an end-to-end manner for {\em 200} epochs, which is the general approach followed in $DNN$-based watermarking techniques~\cite{ref4,hiddn,fang2022pimog,lacuna4}. We used  $MSE$, $LPIPS$~\cite{lpips}, and $L_2$ residual regularization loss functions between the cover image and the watermarked image for training the surrogate encoders. While training surrogate decoders for $HiDDeN$, $ReDMark$, and $PIMoG$, where a bit-string is used as the watermark, we used $BCE$ to calculate the loss between the extracted and the original watermarks. In the same line, $MSE$ and $LPIPS$ loss is used to train the surrogate decoder for {\em Hiding Images in an Image}.

\subsection{Fine-tuning Surrogate Decoder}

In order to demonstrate the efficacy of our attack (Algo \ref{algo-crafting-adversarial-samples}), we will next show that it can successfully adapt to different watermarking techniques through a set of experiments. Before going into the details of our attack results, we briefly discuss our fine-tuning setup. For the fine-tuning, we collected $500$ watermarked images and their watermarks from each of the four watermarking techniques. In the case of $HiDDeN$, $ReDMark$ and $PIMoG$,  each watermarked image is embedded with a unique watermark generated from randomly sampled bits, whereas for {\em Hiding Images in an Image}, we use randomly sampled images from the $Imagenet$ dataset as the watermark. Subsequently, the four instances of the trained surrogate decoder are fine-tuned on the watermarked images of the respective target decoder. The results show that fine-tuning the surrogate decoder for $100$ epochs is sufficient to attack the target decoder successfully.

\subsection{Attack Validation}
\label{sett}

 In order to validate our attack, we generate $10000$ watermarked images using each of the four target watermarking techniques. Using each of these watermarked images, we attack the corresponding fine-tuned surrogate decoder using Algo~\ref{algo-crafting-adversarial-samples} to generate corresponding well-crafted perturbations.  These perturbations are added to the watermarked images and are used to attack the target decoder to evaluate the success rate of our attack.  We tried our attack on the surrogate decoder with $L_1$ and $MSE$ loss (Line $06$, Algo~\ref{algo-crafting-adversarial-samples}). Finally, we chose $MSE$, as the perturbation generated was imperceptible and the attack converged quickly. The initial perturbation $\delta$ is initialized as a zero-filled vector, whereas  $\epsilon$ is chosen as  $-0.3\le \epsilon \le 0.3$. We employ the $Adam$ optimizer with an initial learning rate of $0.001$. The attack converges around $5000$ iterations for all four watermarking techniques.

\subsection{Evaluation}
After the initial training of all four surrogate decoders has an accuracy of more than $90\%$ in successfully extracting the embedded watermark when validated on the test set.  This shows that all four surrogate models have converged successfully. Subsequently, these surrogate decoders are successfully fine-tuned with $500$  watermarked images and their watermarks within $100$ epochs to launch a white-box attack on the surrogate decoder to get the well-crafted perturbation that will fail the target decoder. An experimental analysis is performed for each instance of the surrogate decoder to check if it can be fine-tuned in less than $100$ epochs and also with less than $500$ watermarked images and their watermarks.  The optimum epochs and required watermarked images are shown in $2^{nd}$ and $3^{rd}$ columns of Table~\ref{tab:att}.


\begin{table*}[ht]
\centering
\caption{ Optimal epoch and watermarked image required for fine-tuning different surrogate models along with the image quality of the watermarked images after attacking and adding the well-crafted perturbation using $DLOVE$ attack.  For $PSNR$ and $SSIM$, higher is better. For  $LPIPS$ and $MSE$, lower is better. $ASR$ represent the rate of success on attacking $10000$ watermarked images generated from each $DNN$-based watermarking technique. }

\begin{tabular}{|c|cc|c|ccccc|}
\hline
\textbf{Technique}                                                        & \multicolumn{2}{c|}{\textbf{Fine-Tuning}}                                 & \textbf{Pert Limit ($\epsilon$)} & \multicolumn{5}{c|}{\textbf{Evaluation Matrix}}                                                                                                                  \\ \hline
\multicolumn{1}{|l|}{\textbf{}}                                           & \multicolumn{1}{l|}{\textbf{Epoch}} & \multicolumn{1}{l|}{\textbf{Image}} & \multicolumn{1}{l|}{\textbf{}}  & \multicolumn{1}{c|}{\textbf{PSNR}} & \multicolumn{1}{c|}{\textbf{SSIM}} & \multicolumn{1}{c|}{\textbf{LPIPS}} & \multicolumn{1}{c|}{\textbf{MSE}} & \textbf{ASR} \\ \hline
{\em ReDMark}                                                          & \multicolumn{1}{c|}{40}             & 200                                 & 0.002                           & \multicolumn{1}{c|}{41}            & \multicolumn{1}{c|}{0.97}          & \multicolumn{1}{c|}{0.08}          & \multicolumn{1}{c|}{0.05}       & 98          \\ \hline
{\em HiDDeN}                                                       & \multicolumn{1}{c|}{60}             & 300                                 & 0.008                             & \multicolumn{1}{c|}{38}            & \multicolumn{1}{c|}{0.99}          & \multicolumn{1}{c|}{0.07}           & \multicolumn{1}{c|}{0.15}         & 96          \\ \hline
{\em PIMoG}                                                       & \multicolumn{1}{c|}{70}             & 400                                 & 0.02                           & \multicolumn{1}{c|}{37}            & \multicolumn{1}{c|}{0.99}          & \multicolumn{1}{c|}{0.1}         & \multicolumn{1}{c|}{0.27}       & 93          \\ \hline

{\begin{tabular}[c]{@{}c@{}}{\em Hiding Images}\\ {\em in an Image}\end{tabular}} & \multicolumn{1}{c|}{90}             & 500                                 & 0.1                        & \multicolumn{1}{c|}{33}            & \multicolumn{1}{c|}{0.95}          & \multicolumn{1}{c|}{0.12}          & \multicolumn{1}{c|}{0.36}         & 89           \\ \hline
\end{tabular}
\vspace*{3mm}
\label{tab:att} 
\end{table*}

\vspace{-2em}

 \begin{figure}[!htb]
\begin{minipage}[h]{0.15\linewidth}
\centering
\includegraphics[width=\linewidth]{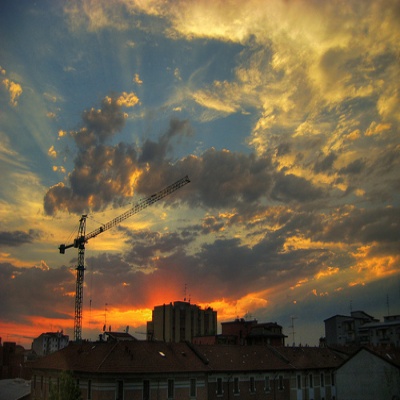}
 Normal \\ Image
\end{minipage}%
\hfill\vrule\hfill
\begin{minipage}[h]{0.15\linewidth}
\centering
\includegraphics[width=\linewidth]{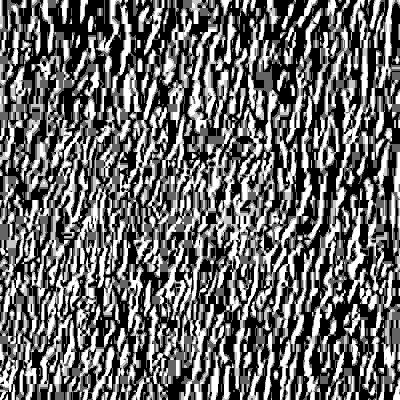}
Added \\ Perturbation 
\end{minipage}%
\hfill\vrule\hfill
\begin{minipage}[h]{0.15\linewidth}
\centering
\includegraphics[width=\linewidth]{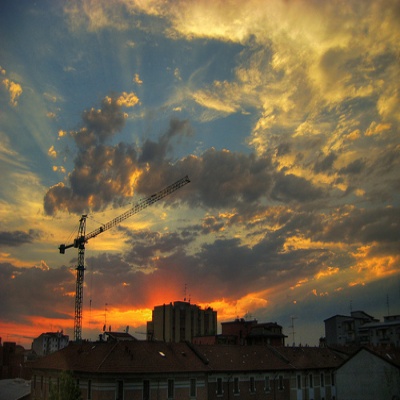}
 Attacked \\ Image
\end{minipage}%
\hfill\vrule\hfill
\begin{minipage}[h]{0.15\linewidth}
\centering
\includegraphics[width=\linewidth]{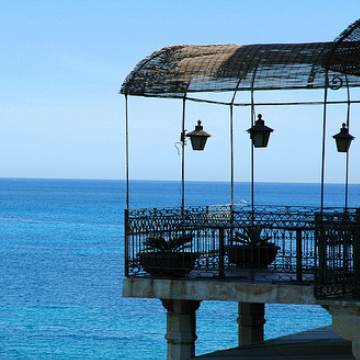}
 Normal \\ Image
\end{minipage}%
\hfill\vrule\hfill
\begin{minipage}[h]{0.15\linewidth}
\centering
\includegraphics[width=\linewidth]{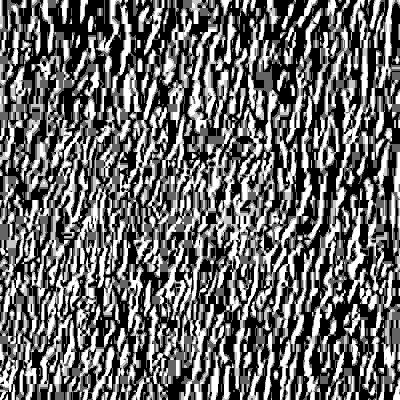}
Added \\ Perturbation 
\end{minipage}%
\hfill\vrule\hfill
\begin{minipage}[h]{0.15\linewidth}
\centering
\includegraphics[width=\linewidth]{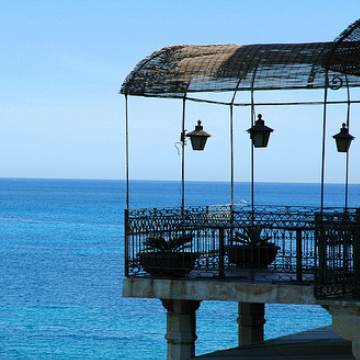}
 Attacked \\ Image
\end{minipage}%
\vspace*{3mm}
\caption{ The well-crafted imperceptible perturbation is successfully added to the original watermarked image without deteriorating the image quality of the watermarked image.}
\label{steg}
\end{figure}

The evaluations are made using the settings mentioned in the Section~\ref{sett}. One of the most important metrics in our evaluations is the Attack Success Rate ($ASR$), which defines the percentage of adversarial examples that lead to successful attacks in the target decoder. Furthermore,  in order to evaluate the quality and the similarity of the attacked (perturbed) watermarked images in comparison to their respective original watermarked images, we use $MSE$, Peak Signal-to-Noise Ratio ($PSNR$), Structural Similarity Index ($SSIM$),  and  $LPIPS$ (from Alexnet Network). These image quality metrics are used in combination with visual analysis to show the quality of our $DLOVE$, $i.e.$, our watermark overwriting attack. In our terms, a good quality attack not only fails the target decoder without leaving any visual traces (artifacts) in the attacked watermarked image. In this line, the $MSE$ and $PSNR$ metrics provide pixel-wise error measurement, which helps the difference between the attacked watermarked images and their respective original watermarked images with respect to the pixels and their orientations. At the same time, the $SSIM$ and $LPIPS$ metrics measure the image quality specifically such that there is no degradation of an image by adding perturbation. 

\vspace{-1em}

\begin{figure}[!htb]
\begin{minipage}[h]{0.18\linewidth}
\centering
\includegraphics[width=\linewidth]{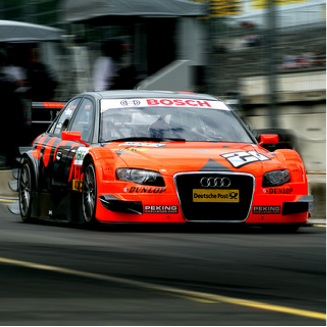}
 Normal \\ Image
\end{minipage}%
\hfill\vrule\hfill
\begin{minipage}[h]{0.18\linewidth}
\centering
\includegraphics[width=\linewidth]{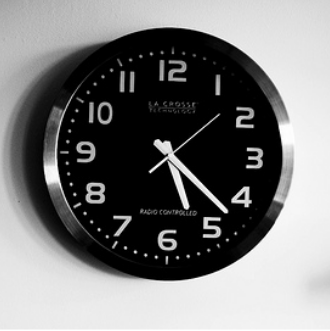}
 Embedded \\ Watermark
\end{minipage}%
\hfill\vrule\hfill
\begin{minipage}[h]{0.18\linewidth}
\centering
\includegraphics[width=\linewidth]{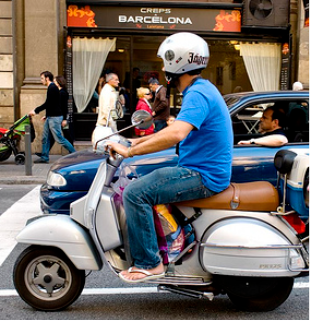}
Target \\ Watermark 
\end{minipage}%
\hfill\vrule\hfill
\begin{minipage}[h]{0.18\linewidth}
\centering
\includegraphics[width=\linewidth]{img/img1.png}
 Attacked \\ Image
\end{minipage}%
\hfill\vrule\hfill
\begin{minipage}[h]{0.18\linewidth}
\centering
\includegraphics[width=\linewidth]{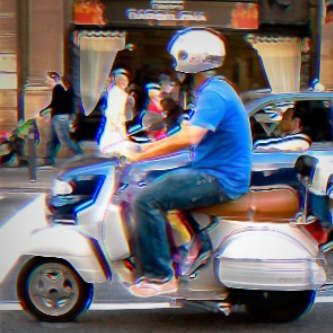}
 Extracted \\ Watermark
\end{minipage}%
\vspace*{3mm}
\caption{ Result of attacking the watermarked image created by the technique of {\em Hiding Images in an Image} using the $DLOVE$ attack.}
\label{water}
\end{figure}

\vspace{-1.5em}

\begin{figure}[!htb]
\begin{minipage}[h]{0.15\linewidth}
\centering
\includegraphics[width=\linewidth]{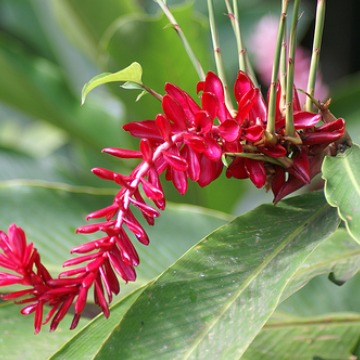}
 Normal \\ Image
\end{minipage}%
\hfill\vrule\hfill
\begin{minipage}[h]{0.15\linewidth}
\centering
\includegraphics[width=\linewidth]{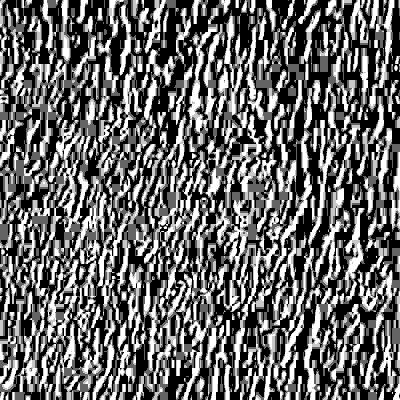}
 Added \\ Perturbation 
\end{minipage}%
\hfill\vrule\hfill
\begin{minipage}[h]{0.15\linewidth}
\centering
\includegraphics[width=\linewidth]{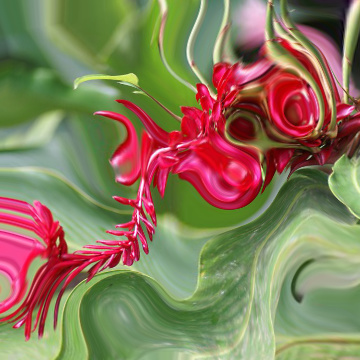}
 Attacked \\ Image
\end{minipage}%
\hfill\vrule\hfill
\begin{minipage}[h]{0.15\linewidth}
\centering
\includegraphics[width=\linewidth]{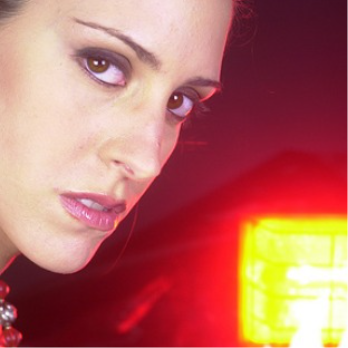}
 Normal \\ Image
\end{minipage}%
\hfill\vrule\hfill
\begin{minipage}[h]{0.15\linewidth}
\centering
\includegraphics[width=\linewidth]{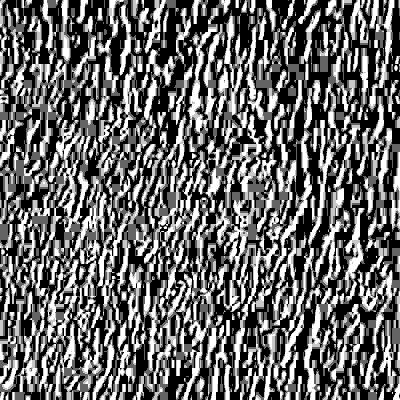}
 Added \\ Perturbation 
\end{minipage}%
\hfill\vrule\hfill
\begin{minipage}[h]{0.15\linewidth}
\centering
\includegraphics[width=\linewidth]{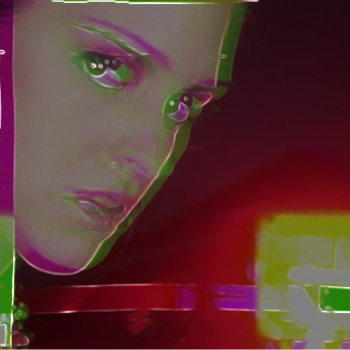}
 Attacked \\ Image
\end{minipage}%
\vspace*{2mm}
\caption{Artifacts appear on attacking some watermarked images using $DLOVE$ when cosine similarity is less than $0.1$.}
\label{arti}
\end{figure}

The performance of our $DLOVE$ attack on different techniques is shown in Table~\ref{tab:att}. Figure~\ref{steg}  show the watermarked image quality after the $DLOVE$ attack where the normal image refers to the original watermarked image and the attacked image refers to the perturbed watermarked image after $DLOVE$ attack. Figure~\ref{water} shows the quality of the target watermark extracted after attacking the technique of {\em Hiding Images in an Image} using $DLOVE$ attack.   The attacked images maintained low values of $MSE$, $LPIPS$, and high values of  $PSNR$ (in \%) and $SSIM$ scores, indicating the added perturbation's imperceptibility. The $ASR$ is $90\%$ for almost all the techniques, highlighting the efficacy of the $DLOVE$ attack. 
Our attack fails for instances where the cosine similarity between the original and target watermarks is less than $0.1$. In order to get a $100\%$ success rate in these cases, we had to sacrifice in perturbation limit, which is increased to $0.5$. This took a toll on all other metrics. Thus, the $MSE$ and $LPIPS$ increased, while $PSNR$ and $SSIM$ decreased significantly. Noticeable artifacts also appeared in the attacked images, as shown in Figure~\ref{arti}.  

\vspace{-1em}

\vspace{-3em}

\subsection{Discussion}

The initial training of the surrogate model for {\em ReDMark} and $HiDDeN$ have a good accuracy and fine-tuning with the target watermarked image leads to a successful $DLOVE$ attack. There were some problems with {\em Hiding Images in an Image} and {\em PIMoG} where the initial surrogate model had good accuracy and similar fine-tuning was performed, but still the $DLOVE$ attack was unsuccessful.  In the case of  {\em Hiding Images in an Image} after the $DLOVE$ attack, the target decoder could recover a distorted watermark that neither matched with the original watermark nor matched with the target watermark. To overcome this issue, we have added $LPIPS$ loss (from $VGG$-$19$ Network) along with the initial $MSE$ loss while training the surrogate decoder with the surrogate dataset. This led to a successful $DLOVE$ attack when we fine-tuned the surrogate model for $90$ epochs with $500$ watermarked image of the target decoder.  In the case of, {\em PiMoG}, the target decoder recovers the original watermark successfully, even with the perturbed watermarked image.   This was possible due to the presence of screen shooting robustness. We introduced perspective warp, motion blur, and colour manipulations to overcome the issue while initially training our surrogate model. Subsequently,  the surrogate decoder is fine-tuned for $70$ epochs with $400$ watermarked images of the target decoder, which led to a successful $DLOVE$ attack. This shows that the $DLOVE$ attack needs minute tweaking in surrogate training and fine-tuning such that it can be adaptable to different techniques. 
 \vspace{-1em}

\section{Conclusion}

In this work, we have proposed the $DLOVE$ attack on  $DNN$-based watermarking techniques by leveraging adversarial machine learning.  The attack shows that modern $DNN$-based watermarking techniques are vulnerable to the $DLOVE$ attack. The proposed $DLOVE$ attack raises a clear question on the security of the existing $DNN$-based watermarking techniques. It provides a new attack vector to the designer community to assess the security of their $DNN$-based watermarking techniques.


%
%
\bibliographystyle{splncs04}
\bibliography{main}

\begin{thebibliography}{10}
\providecommand{\url}[1]{\texttt{#1}}
\providecommand{\urlprefix}{URL }
\providecommand{\doi}[1]{https://doi.org/#1}

\bibitem{realapp3}
Adobe on watermarking ai-generated photos. \url{https://blog.adobe.com/en/publish/2023/10/10/new-content-credentials-icon-transparency}, accessed: 2022-20-02

\bibitem{lacuna4}
Ahmadi, M., Norouzi, A., Karimi, N., Samavi, S., Emami, A.: Redmark: Framework for residual diffusion watermarking based on deep networks. Expert Systems with Applications  \textbf{146},  113157 (2020)

\bibitem{sit}
Atito, S., Awais, M., Kittler, J.: Sit: Self-supervised vision transformer. arXiv preprint arXiv:2104.03602  (2021)

\bibitem{ref4}
Baluja, S.: Hiding images within images. IEEE transactions on pattern analysis and machine intelligence  \textbf{42}(7),  1685--1697 (2019)

\bibitem{robust}
Berghel, H., O'Gorman, L.: Protecting ownership rights through digital watermarking. Computer  \textbf{29}(7),  101--103 (1996). \doi{10.1109/2.511977}

\bibitem{dest}
Corley, I., Lwowski, J., Hoffman, J.: Destruction of image steganography using generative adversarial networks. arXiv preprint arXiv:1912.10070  (2019)

\bibitem{Coxwatermark}
Cox, I., Kilian, J., Leighton, F., Shamoon, T.: Secure spread spectrum watermarking for multimedia. IEEE Transactions on Image Processing  \textbf{6}(12),  1673--1687 (1997). \doi{10.1109/83.650120}

\bibitem{cox2002digital}
Cox, I., Miller, M., Bloom, J., Honsinger, C.: Digital watermarking. Journal of Electronic Imaging  \textbf{11}(3),  414--414 (2002)

\bibitem{cox2002first}
Cox, I.J., Miller, M.L.: The first 50 years of electronic watermarking. EURASIP Journal on Advances in Signal Processing  \textbf{2002}, ~1--7 (2002)

\bibitem{deng2009imagenet}
Deng, J., Dong, W., Socher, R., Li, L.J., Li, K., Fei-Fei, L.: Imagenet: A large-scale hierarchical image database. In: 2009 IEEE conference on computer vision and pattern recognition. pp. 248--255. Ieee (2009)

\bibitem{fang2020deep}
Fang, H., Chen, D., Huang, Q., Zhang, J., Ma, Z., Zhang, W., Yu, N.: Deep template-based watermarking. IEEE Transactions on Circuits and Systems for Video Technology  \textbf{31}(4),  1436--1451 (2020)

\bibitem{fang2022pimog}
Fang, H., Jia, Z., Ma, Z., Chang, E.C., Zhang, W.: Pimog: An effective screen-shooting noise-layer simulation for deep-learning-based watermarking network. In: Proceedings of the 30th ACM International Conference on Multimedia. pp. 2267--2275 (2022)

\bibitem{harness}
Goodfellow, I.J., Shlens, J., Szegedy, C.: Explaining and harnessing adversarial examples. arXiv preprint arXiv:1412.6572  (2014)

\bibitem{realapp2}
Google on watermarking ai-generated contents. \url{https://deepmind.google/technologies/synthid/}, accessed: 2022-20-02

\bibitem{simba}
Guo, C., Gardner, J., You, Y., Wilson, A.G., Weinberger, K.: Simple black-box adversarial attacks. In: International Conference on Machine Learning. pp. 2484--2493. PMLR (2019)

\bibitem{mir}
Huiskes, M.J., Thomee, B., Lew, M.S.: New trends and ideas in visual concept detection: The mir flickr retrieval evaluation initiative. In: Proceedings of the international conference on Multimedia information retrieval. pp. 527--536 (2010)

\bibitem{limba}
Ilyas, A., Engstrom, L., Athalye, A., Lin, J.: Black-box adversarial attacks with limited queries and information. In: International conference on machine learning. pp. 2137--2146. PMLR (2018)

\bibitem{spatial}
Jaderberg, M., Simonyan, K., Zisserman, A., et~al.: Spatial transformer networks. Advances in neural information processing systems  \textbf{28} (2015)

\bibitem{lacuna6}
Jia, Z., Fang, H., Zhang, W.: Mbrs: Enhancing robustness of dnn-based watermarking by mini-batch of real and simulated jpeg compression. In: Proceedings of the 29th ACM international conference on multimedia. pp. 41--49 (2021)

\bibitem{pixel}
Jung, D., Bae, H., Choi, H.S., Yoon, S.: Pixelsteganalysis: Pixel-wise hidden information removal with low visual degradation. IEEE Transactions on Dependable and Secure Computing  (2021)

\bibitem{deepwatermark1}
Kandi, H., Mishra, D., Gorthi, S.R.S.: Exploring the learning capabilities of convolutional neural networks for robust image watermarking. Computers \& Security  \textbf{65},  247--268 (2017)

\bibitem{cifar}
Krizhevsky, A., Hinton, G., et~al.: Learning multiple layers of features from tiny images  (2009)

\bibitem{lin2014microsoft}
Lin, T.Y., Maire, M., Belongie, S., Hays, J., Perona, P., Ramanan, D., Doll{\'a}r, P., Zitnick, C.L.: Microsoft coco: Common objects in context. In: Computer Vision--ECCV 2014: 13th European Conference, Zurich, Switzerland, September 6-12, 2014, Proceedings, Part V 13. pp. 740--755. Springer (2014)

\bibitem{liu2023erase}
Liu, H., Xiang, T., Guo, S., Li, H., Zhang, T., Liao, X.: Erase and repair: An efficient box-free removal attack on high-capacity deep hiding. IEEE Transactions on Information Forensics and Security  (2023)

\bibitem{survey2}
Liu, T., Qiu, Z.d.: The survey of digital watermarking-based image authentication techniques. In: 6th International Conference on Signal Processing, 2002. vol.~2, pp. 1556--1559. IEEE (2002)

\bibitem{lacuna1}
Liu, Y., Guo, M., Zhang, J., Zhu, Y., Xie, X.: A novel two-stage separable deep learning framework for practical blind watermarking. In: Proceedings of the 27th ACM International conference on multimedia. pp. 1509--1517 (2019)

\bibitem{liu2016delving}
Liu, Y., Chen, X., Liu, C., Song, D.: Delving into transferable adversarial examples and black-box attacks. arXiv preprint arXiv:1611.02770  (2016)

\bibitem{robust1}
Lu, C.S., Huang, S.K., Sze, C.J., Liao, H.Y.M.: Cocktail watermarking for digital image protection. IEEE Transactions on Multimedia  \textbf{2}(4),  209--224 (2000). \doi{10.1109/6046.890056}

\bibitem{951542}
Lu, C.S., Liao, H.Y.: Multipurpose watermarking for image authentication and protection. IEEE Transactions on Image Processing  \textbf{10}(10),  1579--1592 (2001). \doi{10.1109/83.951542}

\bibitem{lacuna2}
Luo, X., Zhan, R., Chang, H., Yang, F., Milanfar, P.: Distortion agnostic deep watermarking. In: Proceedings of the IEEE/CVF conference on computer vision and pattern recognition. pp. 13548--13557 (2020)

\bibitem{towards}
Madry, A., Makelov, A., Schmidt, L., Tsipras, D., Vladu, A.: Towards deep learning models resistant to adversarial attacks. arXiv preprint arXiv:1706.06083  (2017)

\bibitem{realapp1}
Meta on watermark ai-generated photos. \url{https://about.fb.com/news/2024/02/labeling-ai-generated-images-on-facebook-instagram-and-threads/}, accessed: 2022-20-02

\bibitem{deepfool}
Moosavi-Dezfooli, S.M., Fawzi, A., Frossard, P.: Deepfool: a simple and accurate method to fool deep neural networks. In: Proceedings of the IEEE conference on computer vision and pattern recognition. pp. 2574--2582 (2016)

\bibitem{nguyen2015deep}
Nguyen, A., Yosinski, J., Clune, J.: Deep neural networks are easily fooled: High confidence predictions for unrecognizable images. In: Proceedings of the IEEE conference on computer vision and pattern recognition. pp. 427--436 (2015)

\bibitem{papernot2016transferability}
Papernot, N., McDaniel, P., Goodfellow, I.: Transferability in machine learning: from phenomena to black-box attacks using adversarial samples. arXiv preprint arXiv:1605.07277  (2016)

\bibitem{papernot2017practical}
Papernot, N., McDaniel, P., Goodfellow, I., Jha, S., Celik, Z.B., Swami, A.: Practical black-box attacks against machine learning. In: Proceedings of the 2017 ACM on Asia conference on computer and communications security. pp. 506--519 (2017)

\bibitem{papernot2016limitations}
Papernot, N., McDaniel, P., Jha, S., Fredrikson, M., Celik, Z.B., Swami, A.: The limitations of deep learning in adversarial settings. In: 2016 IEEE European symposium on security and privacy (EuroS\&P). pp. 372--387. IEEE (2016)

\bibitem{pibre2015deep}
Pibre, L., J{\'e}r{\^o}me, P., Ienco, D., Chaumont, M.: Deep learning is a good steganalysis tool when embedding key is reused for different images, even if there is a cover source-mismatch. arXiv preprint arXiv:1511.04855  (2015)

\bibitem{qian2015deep}
Qian, Y., Dong, J., Wang, W., Tan, T.: Deep learning for steganalysis via convolutional neural networks. In: Media Watermarking, Security, and Forensics 2015. vol.~9409, pp. 171--180. SPIE (2015)

\bibitem{raj2021survey}
Raj, N.N., Shreelekshmi, R.: A survey on fragile watermarking based image authentication schemes. Multimedia Tools and Applications  \textbf{80},  19307--19333 (2021)

\bibitem{unet}
Ronneberger, O., Fischer, P., Brox, T.: U-net: Convolutional networks for biomedical image segmentation. In: Medical Image Computing and Computer-Assisted Intervention--MICCAI 2015: 18th International Conference, Munich, Germany, October 5-9, 2015, Proceedings, Part III 18. pp. 234--241. Springer (2015)

\bibitem{shaik2022review}
Shaik, A.S., Karsh, R.K., Islam, M., Laskar, R.H.: A review of hashing based image authentication techniques. Multimedia Tools and Applications pp. 1--28 (2022)

\bibitem{intriguing}
Szegedy, C., Zaremba, W., Sutskever, I., Bruna, J., Erhan, D., Goodfellow, I., Fergus, R.: Intriguing properties of neural networks. arXiv preprint arXiv:1312.6199  (2013)

\bibitem{lacuna5}
Vukoti{\'c}, V., Chappelier, V., Furon, T.: Are deep neural networks good for blind image watermarking? In: 2018 IEEE International Workshop on Information Forensics and Security (WIFS). pp.~1--7. IEEE (2018)

\bibitem{wang2002wavelet}
Wang, Y., Doherty, J.F., Van~Dyck, R.E.: A wavelet-based watermarking algorithm for ownership verification of digital images. IEEE transactions on image processing  \textbf{11}(2),  77--88 (2002)

\bibitem{realapp4}
Single-frame \& image forensic watermarking. \url{https://castlabs.com/image-watermarking/}, accessed: 2022-20-02

\bibitem{Wong1998APK}
Wong, P.W.: A public key watermark for image verification and authentication. Proceedings 1998 International Conference on Image Processing. ICIP98 (Cat. No.98CB36269)  \textbf{1},  455--459 vol.1 (1998), \url{https://api.semanticscholar.org/CorpusID:15447332}

\bibitem{ping}
Wong, P.W., Memon, N.: Secret and public key image watermarking schemes for image authentication and ownership verification. IEEE Transactions on Image Processing  \textbf{10}(10),  1593--1601 (2001). \doi{10.1109/83.951543}

\bibitem{lpips}
Zhang, R., Isola, P., Efros, A.A., Shechtman, E., Wang, O.: The unreasonable effectiveness of deep features as a perceptual metric. In: Proceedings of the IEEE conference on computer vision and pattern recognition. pp. 586--595 (2018)

\bibitem{4303093}
Zhang, X., Wang, S.: Statistical fragile watermarking capable of locating individual tampered pixels. IEEE Signal Processing Letters  \textbf{14}(10),  727--730 (2007). \doi{10.1109/LSP.2007.896436}

\bibitem{deepwatermark2}
Zhong, X., Huang, P.C., Mastorakis, S., Shih, F.Y.: An automated and robust image watermarking scheme based on deep neural networks. IEEE Transactions on Multimedia  \textbf{23},  1951--1961 (2020)

\bibitem{zhou2018transferable}
Zhou, W., Hou, X., Chen, Y., Tang, M., Huang, X., Gan, X., Yang, Y.: Transferable adversarial perturbations. In: Proceedings of the European Conference on Computer Vision (ECCV). pp. 452--467 (2018)

\bibitem{hiddn}
Zhu, J., Kaplan, R., Johnson, J., Fei-Fei, L.: Hidden: Hiding data with deep networks. In: Proceedings of the European conference on computer vision (ECCV). pp. 657--672 (2018)

\end{thebibliography}
\end{document}


\title{ $DLOVE$: A new Security Evaluation Tool for Deep Learning Based Watermarking Techniques.
Appendix}

\maketitle

\section{Alternate Setup of Surrogate Model}
In this setup, we want to find if we can make a common surrogate model that is trained and fine-tuned only once and is able to attack multiple techniques. Each technique mentioned above uses a different resolution of the cover image and watermark size. As $HiDDeN$, $ReDMark$, and $PIMoG$ use bit-string as the watermark, we have used one surrogate model (encoder and decoder),  for these techniques. The size of the cover image, watermarked image and watermark of the surrogate model is fixed at $224 \times 224 \times 3$, $224 \times 224 \times 3$ and $10$ bits, respectively.   We have used $UNet$~\cite{unet} architecture for the surrogate encoder, whereas, for the surrogate decoder, we have used is spatial transformer~\cite{spatial} with seven convolutional layers followed by two fully connected layers. We have used the $Mirflickr$~\cite{mir} dataset as our surrogate dataset, consisting of one million images with varied contexts, lighting, and themes, from the social photography site $Flickr$. We used $50$k images in our training set and $10$k in our test set. We trained the surrogate model in an end-to-end manner for {\em 200} epochs, which is the general approach followed in $DNN$-based watermarking techniques~\cite{ref4,hiddn,fang2022pimog, lacuna4}. We used  $MSE$, $LPIPS$~\cite{lpips}, and $L_2$ residual regularization loss functions between the cover image and the watermarked image for training the surrogate encoders. While training the surrogate decoder where a bit-string of $!0$-bit is used as the watermark, we used $BCE$ to calculate the loss between the extracted and the original watermarks.  We collected $500$ watermarked images and their watermarks from each for $HiDDeN$, $ReDMark$ and $PIMoG$ for the fine-tuning. The watermarked images are upscaled to $224 \times 224 \times 3$.  In the case of $HiDDeN$, $ReDMark$ and $PIMoG$,  each watermarked image is embedded with a unique watermark generated from randomly sampled bits which is capped at $10$ bits for this setup. Subsequently, the trained surrogate decoder is fine-tuned on the $1500$ watermarked images collected from  $HiDDeN$, $ReDMark$, and $PIMoG$. This fine-tuned decoder can attack the techniques of  $HiDDeN$, $ReDMark$ and $PIMoG$ at once.

\subsection{Attack Validation}
\label{sett}

 In order to validate our attack, we generate $1000$ watermarked images using each one of the techniques of  $HiDDeN$, $ReDMark$, and $PIMoG$. Using each of these watermarked images, we attack the fine-tuned surrogate decoder to generate corresponding well-crafted perturbations. To attack the fine-tuned surrogate decoder, the watermarked image is upscaled to $224 \times 224 \times 3$ and the well-crafted perturbation is generated accordingly. Once the attack is successful on the surrogate decoder, the attacked image is downscaled to the resolution of the target decoder to verify the success.   We chose $MSE$ as the loss function while generating the perturbation. The initial perturbation $\delta$ is initialized as a zero-filled vector, whereas  $\epsilon$ is chosen as  $-0.3\le \epsilon \le 0.3$. We employ the $Adam$ optimizer with an initial learning rate of $0.001$. The attack converges around $8000$ iterations for all four watermarking techniques.

\subsection{Evaluation}
After the initial training of the surrogate decoder has an accuracy of more than $90\%$ in successfully extracting the embedded watermark when validated on the test set; this shows that the surrogate model has converged successfully. Subsequently, these surrogate decoder is successfully fine-tuned with $1500$  watermarked images generated from techniques of $HiDDeN$, $ReDMark$ and $PIMoG$ for $100$ epochs to launch a white-box attack on the surrogate decoder to get the well-crafted perturbation that will fail the target decoder. The results were good but not better than the original setup, where one instance was created for each target model. The result can be found in Table~\ref{tab:att}. The fine-tuning with various watermarked images has led to lower $PSNR$ and $SSIM$ and attack success rates while higher $LPIPS$ and $MSE$. This is due to the fact that the surrogate decoder fails to replicate the behaviour of all the target decoders at once. Still, the surrogate decoder successfully removes the watermark, such that the target decoder fails to decode the watermark in most cases. If the attack success rate is measured as failing the decoder to decode the watermark, then the attack success rate for $HiDDeN$, $ReDMark$ and $PIMoG$ are $89\%$, $86\%$ and $78\%$ respectively, which shows the efficiency of the trained surrogate model.

\begin{table*}[ht]
\centering
\caption{ Optimal epoch and watermarked image required for fine-tuning the surrogate model such that a surrogate model can successfully attack $HiDDeN$, $ReDMark$, and $PIMoG$ when the watermark is of $10$ bits.   The image quality of the watermarked images after attacking and adding the well-crafted perturbation using the $DLOVE$ attack is also present.  For $PSNR$ and $SSIM$, higher is better. For  $LPIPS$ and $MSE$, lower is better. $ASR$ represent the rate of success on attacking $1000$ watermarked images generated from each $DNN$-based watermarking technique. }

\begin{tabular}{|c|cc|c|ccccc|}
\hline
\textbf{Technique}                                                        & \multicolumn{2}{c|}{\textbf{Fine-Tuning}}                                 & \textbf{Pert Limit ($\epsilon$)} & \multicolumn{5}{c|}{\textbf{Evaluation Matrix}}                                                                                                                  \\ \hline
\multicolumn{1}{|l|}{\textbf{}}                                           & \multicolumn{1}{l|}{\textbf{Epoch}} & \multicolumn{1}{l|}{\textbf{Image}} & \multicolumn{1}{l|}{\textbf{}}  & \multicolumn{1}{c|}{\textbf{PSNR}} & \multicolumn{1}{c|}{\textbf{SSIM}} & \multicolumn{1}{c|}{\textbf{LPIPS}} & \multicolumn{1}{c|}{\textbf{MSE}} & \textbf{ASR} \\ \hline
{\em ReDMark}                                                          & \multicolumn{1}{c|}{90}             & 500                                 & 0.01                         & \multicolumn{1}{c|}{36}            & \multicolumn{1}{c|}{0.93}          & \multicolumn{1}{c|}{0.12}          & \multicolumn{1}{c|}{0.1}       & 71          \\ \hline
{\em HiDDeN}                                                       & \multicolumn{1}{c|}{90}             & 500                                 & 0.01                             & \multicolumn{1}{c|}{32}            & \multicolumn{1}{c|}{0.96}          & \multicolumn{1}{c|}{0.19}           & \multicolumn{1}{c|}{0.23}         & 68          \\ \hline
{\em PIMoG}                                                       & \multicolumn{1}{c|}{90}             & 500                                 & 0.01                           & \multicolumn{1}{c|}{29}            & \multicolumn{1}{c|}{0.95}          & \multicolumn{1}{c|}{0.27}         & \multicolumn{1}{c|}{0.45}       & 59         \\ \hline

\end{tabular}
\vspace*{3mm}
\label{tab:att} 
\end{table*}

\vspace{-2em}

\section{Reason for successful $DLOVE$ Attack}
We have already highlighted why our $DLOVE$ attack is successful. Here we are adding few more points for the same. In~\cite{ref4} the author has explored the possibility of an attacker training a network to reveal a watermark without having access to the original network. They investigated a potential attack scenario in which the attacker can fail the target decoder of the watermarking technique by acquiring a few instances of watermarked images and the corresponding watermark for the target technique. Furthermore, $DNN$-based watermarking algorithms demonstrate transferability~\cite{intriguing,papernot2016transferability,papernot2017practical}. This means that the decoder of one model can partially restore the secret image encoded by another model. The attacker can roughly know the secret image and find the original secret image corresponding to the revealed image. Therefore, the surrogate model variant of the black-box scenario is proposed as a threat model for deep watermarking attacks.

\section{Challenges}
The encoder being used in {\em ReDMark}  generate low-resolution images compared to the images used in {\em HiDDeN}, {\em Hiding Images in an Image} and $PIMoG$. In the context of adversarial machine learning, reducing the image size will make the attack more challenging because it will also decrease the number of non-robust features that the attacker can exploit~\cite{AML_Not_Bug}. It may also be the case that certain visible artefacts may appear while attacking images with smaller resolutions. To overcome this while attacking the perturbation range is set lower so there is no artifact in the image.

\section{Additional Results}

\begin{figure}[!htb]
\centering
\includegraphics[width=0.7\linewidth]{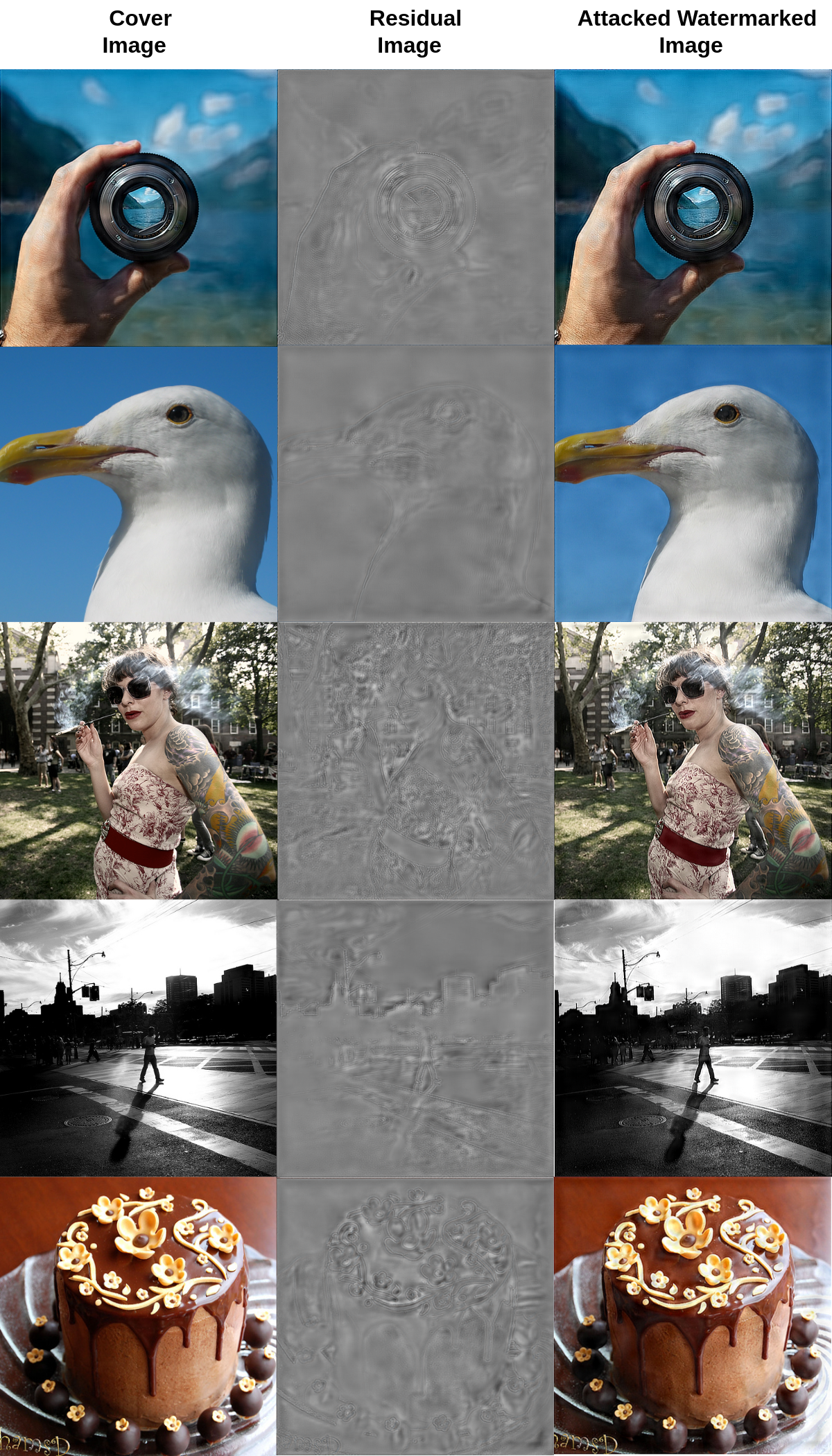}
\caption{Difference between the original cover image and the corresponding attacked image. }
\label{res}
\end{figure}

In this section, we provide more visual details of the watermarked images attacked using $DLOVE$. Figure~\ref{res} shows the residual image, which represents the difference between the cover image and the attacked watermarked image. \textbf{The code is also provided in the supplementary material.} 

\bibliographystyle{splncs04}
\bibliography{main}